\documentclass[11pt]{article}
\usepackage{moriond,epsfig}

\def\Journal#1#2#3#4{{#1} {\bf #2}, #3 (#4)}

\def\MN{\em MNRAS}

\def\be{\begin{equation}}
\def\ee{\end{equation}}
\def\bea{\begin{eqnarray}}
\def\eea{\end{eqnarray}}
\newcommand{\hMpc}{{\ifmmode{h^{-1}{\rm Mpc}}\else{$h^{-1}$Mpc}\fi}}
\newcommand{\hkpc}{{\ifmmode{h^{-1}{\rm kpc}}\else{$h^{-1}$kpc}\fi}}
\newcommand{\hMsun}{{\ifmmode{h^{-1}{\rm {M_{\odot}}}}\else{$h^{-1}{\rm{M_{\odot}}}$}\fi}}
\newcommand{\Msun}{{\ifmmode{{\rm {M_{\odot}}}}\else{${\rm{M_{\odot}}}$}\fi}}

\begin{document}
\vspace*{4cm}
\title{THE MARENOSTRUM UNIVERSE}

\author{S. GOTTL\"OBER$^1$, G. YEPES$^2$, C. WAGNER$^1$, R. SEVILLA$^2$}
\address{$^1$Astrophysical Institute Potsdam, An der Sternwarte 16, 
14482 Potsdam, Germany \\
$^2$ Grupo de Astrof\'\i sica, Universidad Aut\'onoma de Madrid, 
Madrid E-28049 (Spain)}

\maketitle\abstracts { The MareNostrum Universe is one of the biggest
  SPH cosmological simulations done so far. It contains more than 2
  billion particles ($2\times 1024^3$) in a 500 h$^{-1}$ Mpc cubic
  volume.  This simulation has been performed on the MareNostrum
  supercomputer at the Barcelona Supercomputer Center. We have
  obtained more than 0.5 million halos with masses greater than a
  typical Milky Way galaxy halo.  We report results about the halo
  mass function, the shapes of dark matter and gas distributions in
  halos, the baryonic fraction in galaxy clusters and groups, baryon
  oscillations in the dark matter and the halo power spectra as well
  as the distribution and evolution of the gas fraction at large
  scales.  }

\section{Numerical Simulation and Data Analysis}

In our numerical simulation we have assumed the spatially flat
concordance cosmological model with the parameters $\Omega_m = 0.3$,
$\Omega_{bar} = 0.045$, $\Omega_{\Lambda} = 0.7$, the normalization
$\sigma_8 = 0.9$ and the slope $n=1$ of the power spectrum.  Within a
box of $500 \hMpc$ size the linear power spectrum at redshift $z=40$
has been represented by $1024^3$ DM particles of mass $m_{\rm DM} =
8.3 \times 10^{9} h^{-1} \Msun $ and $1024^3$ gas particles of mass
$m_{\rm gas} = 1.5 \times 10^{9} h^{-1} \Msun $. The nonlinear
evolution of structures has been followed by the GADGET II code of V.
Springel \cite{GYWS:vs}.  For the gravitational evolution we
have used the TREEPM algorithm on a homogeneous Eulerian grid
to compute large scale forces by the Particle-Mesh algorithm.  In this
simulation we employed $1024^3$ mesh points to compute the density field
from particle positions and FFT to derive gravitational forces.  Since
the baryonic component is also discretized by the gas particles all
hydrodynamical quantities have to be determined using interpolation
from the gas particles. Within GADGET the equations of gas dynamics
are solved by means of the Smoothed Particle Hydrodynamics method in
its entropy conservation scheme.  To follow structure formation until
redshift $z=0$ we have restricted ourselves to the gas-dynamics
without including dissipative or radiative processes or star
formation.  The spatial force resolution was set to an equivalent
Plummer gravitational softening of $15 \;h^{-1}$ comoving kpc. The SPH
smoothing length was set to the distance to the 40$^{th}$ nearest
neighbor of each SPH particle. In any case, we do not allow smoothing
scales to be smaller than the gravitational softening of the gas
particles. Using for three weeks 512 processors of the MareNostrum
supercomputer at BSC Barcelona (this time corresponds to 29 CPU years)
we have finished the simulation and created the {\em MareNostrum
  Universe}.

\begin{figure}
\begin{center}
\includegraphics[width=7.5cm]{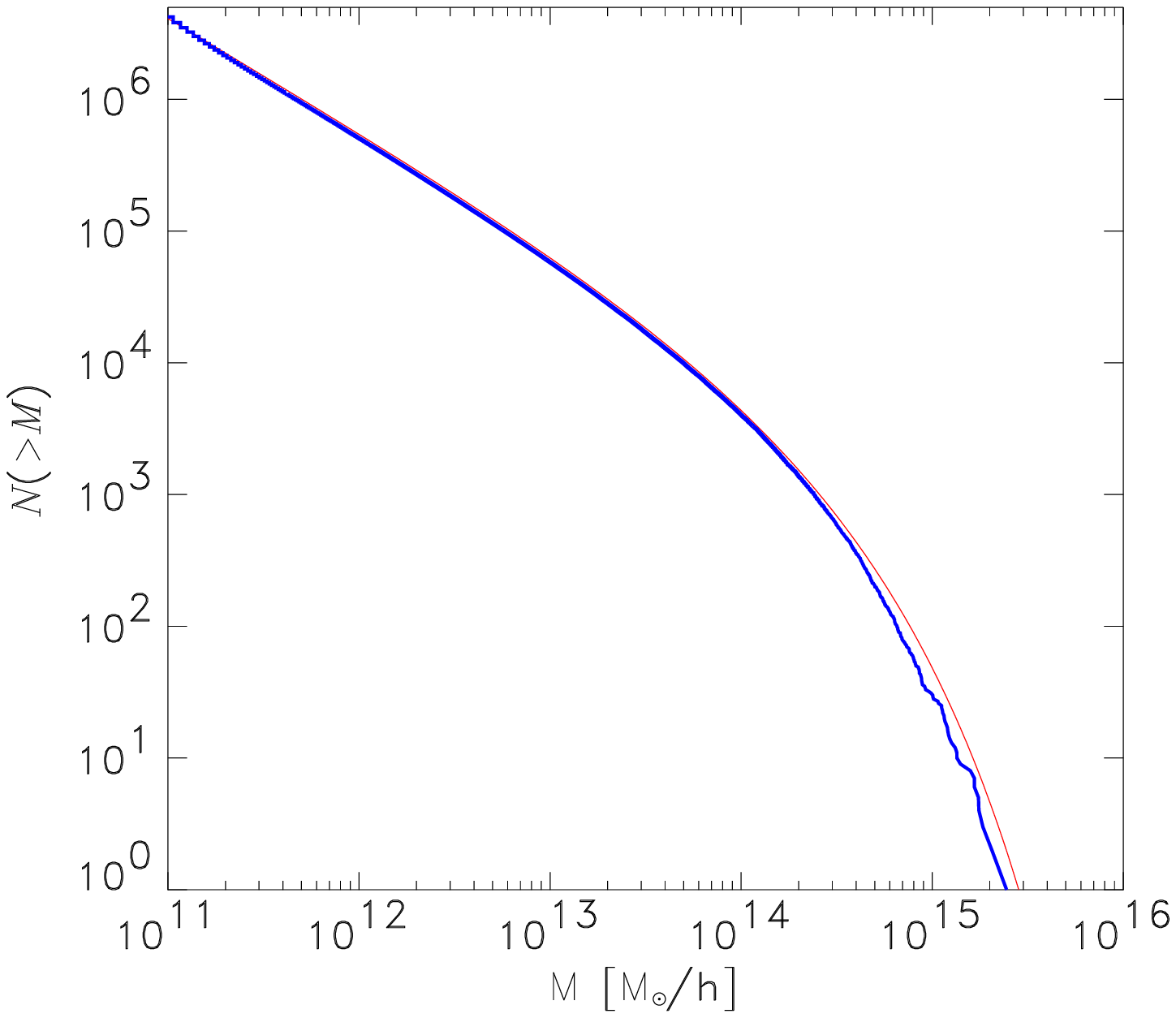}
\includegraphics[width=7.5cm]{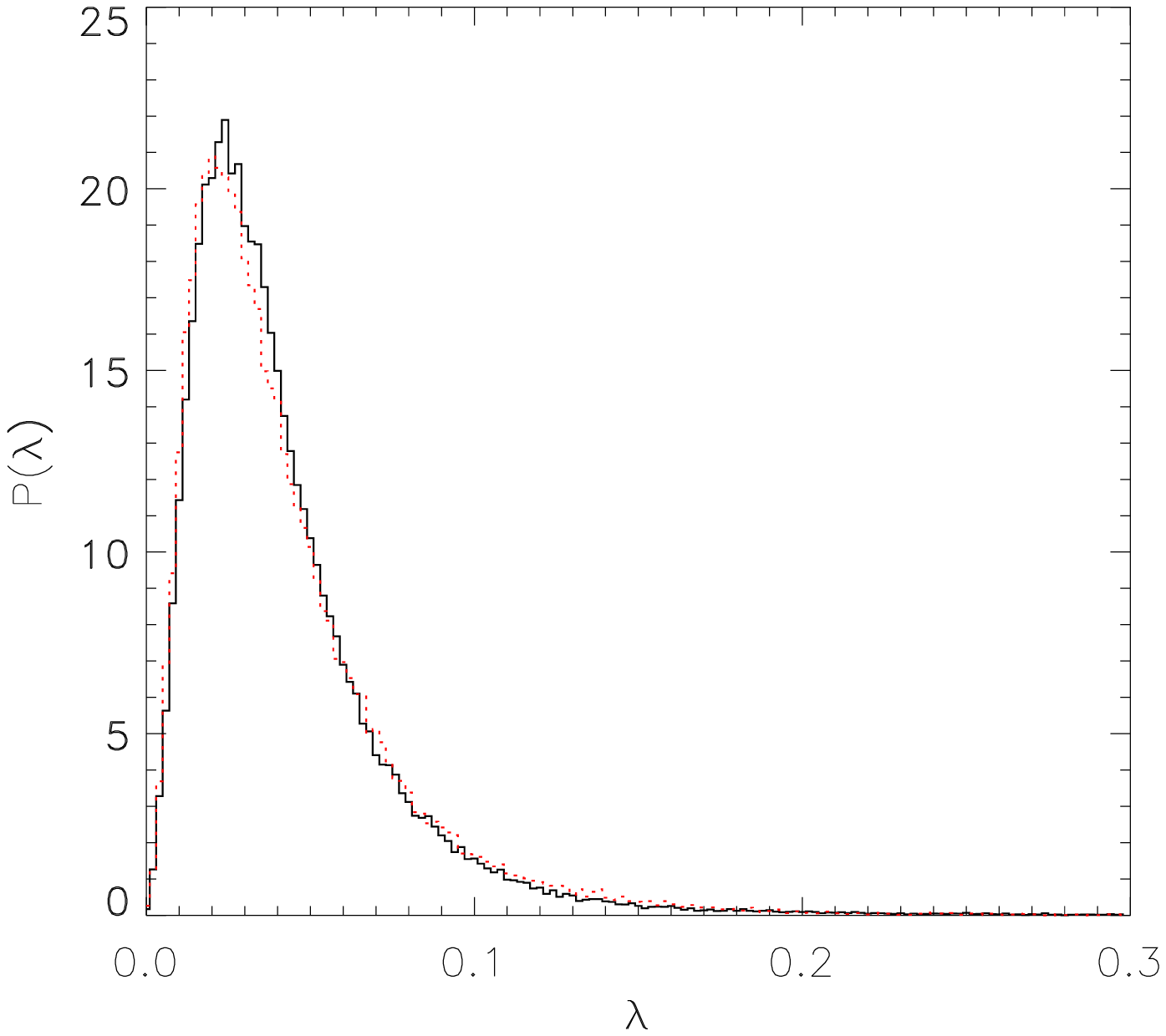}

\caption{Left: Mass function of objects in the MareNostrum Universe, dotted 
 line for Sheth-Tormen approximation. Right: Probability distribution
 of spin parameter of  halos with more than 500 particles for  virial
 overdensity (solid) and 8 times viral overdensity (dotted). }
\label{GYWS:mass_func+spin}
\end{center}
\end{figure}

With 2 billion particles (DM and gas) the code produces 64 Gb of data
per time step. To follow the evolution of structures we have stored
135 time steps equally separated by $10^8$ years.  It is a challenge
to find structures and substructures in the distribution of DM and gas
particles. We have used a parallel hierarchical friends-of-friends
algorithm which is based on the calculation of the minimum spanning
tree of the given particle distribution. The minimum spanning tree of
any point distribution is a unique, well defined quantity which
describes the clustering properties of the point process completely.
Based on the minimum spanning tree we sort the particles in such a way
that we get a particle-cluster ordered sequence. Any object defined at
any linking length is a segment of this sequence. Thus we can easily
extract all objects and their substructures. The minimum spanning tree
and the FOF-analysis have been done on JUMP J\"ulich.

In the left panel of Fig.~\ref{GYWS:mass_func+spin} we show the total
number of objects identified in the simulation with masses ($> M$).
The MareNostrum Universe contains 4060 clusters of galaxies with
masses larger than $10^{14} \hMsun$, more than 58000 groups and
clusters with masses larger than $10^{13} \hMsun$. It contains about
0.5 million objects with masses between $10^{12} \hMsun$ and $10^{13}
\hMsun$. In total we have identified at redshift $z=0$ more than 2
million objects with more than 20 DM particles. The spin parameter
probability function is shown in Fig.~\ref{GYWS:mass_func+spin}
(right) for halos defined at different overdensities. The most likely
value is $ \lambda_0 = 0.023 $, in good agreement with other numerical
studies.  There is no significant difference between the probability
distribution of the spin parameters at virial and 8 times virial
overdensity.

\begin{figure}[t]
\begin{center}
\includegraphics[width=7.5cm]{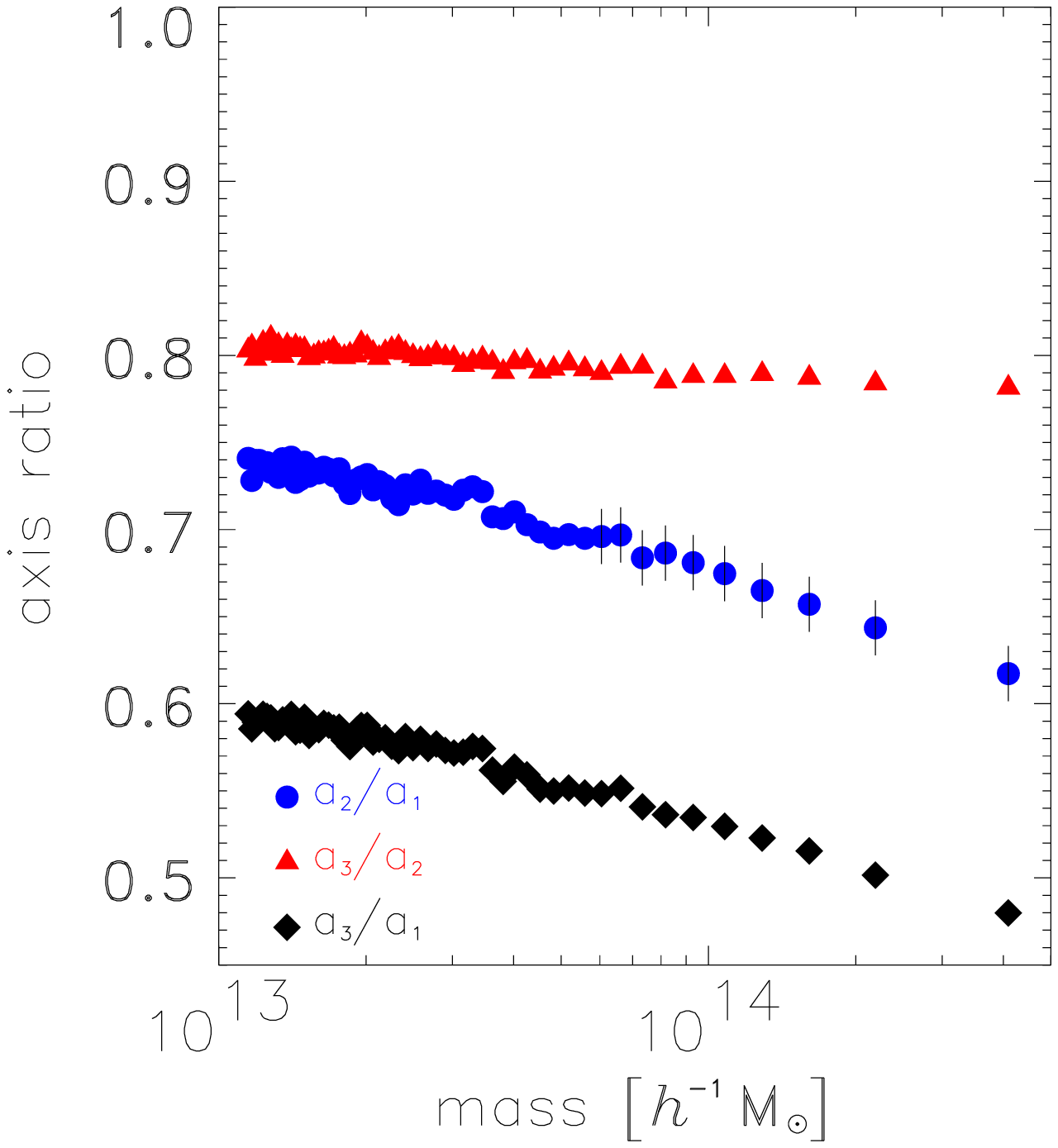}
\includegraphics[width=7.5cm]{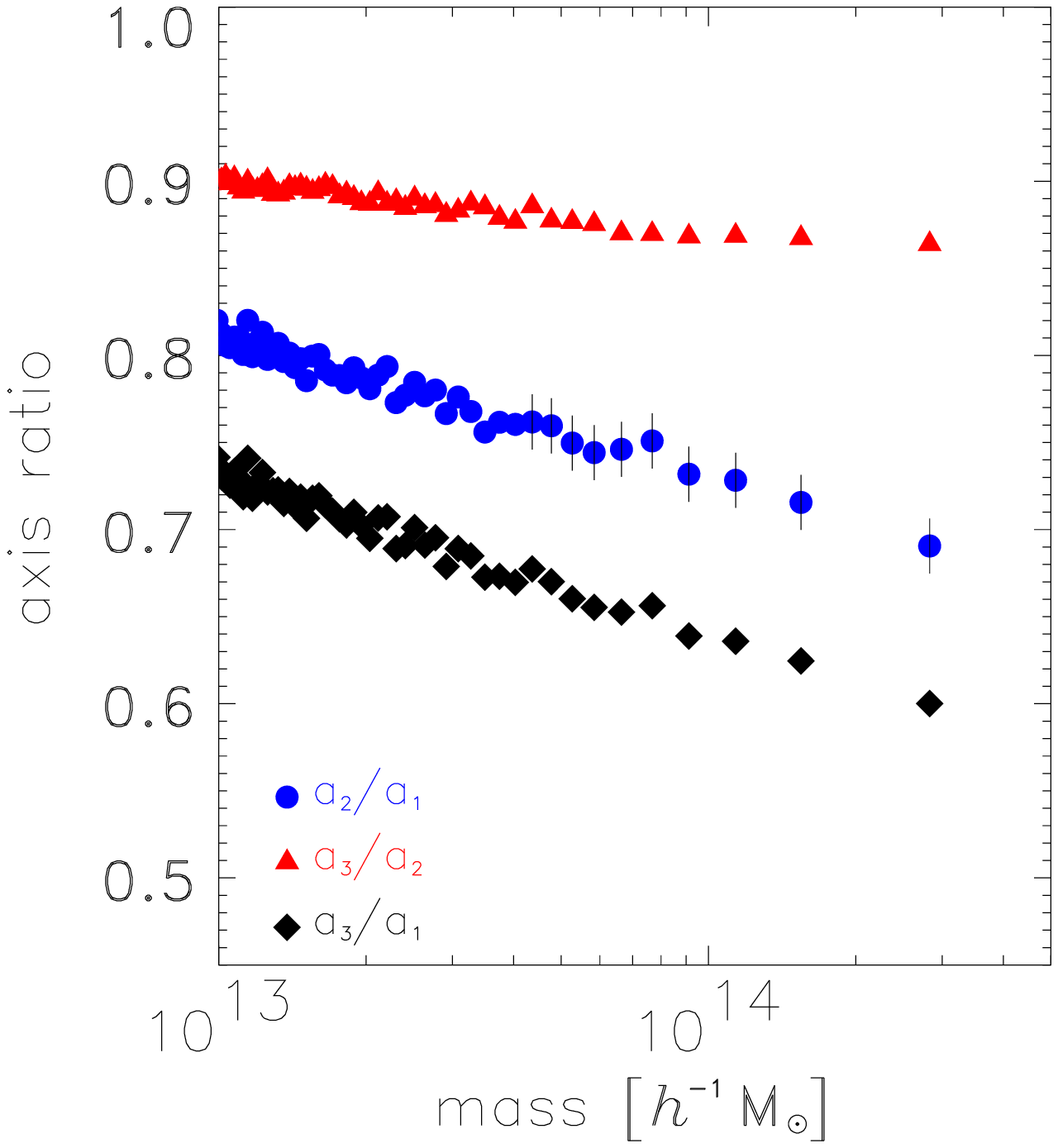}
\caption{Left: The ratio between the main axes of the DM 
  halos of clusters and groups depending on their mass. Right: The
  same for the gas distribution.  }
\label{GYWS:shape}
\end{center}
\end{figure}

The shape of the objects can be described by three-axial ellipsoids
with main axes $a_1 \geq a_2 \geq a_3$. In Fig.~\ref{GYWS:shape} we
show the shape of group and cluster sized objects depending on mass.
Filled circles denote the mean ratio of $a_2/a_1$ of 1000 objects in
each bin. (The bars on the 10 most massive objects denote the
corresponding Poisson errors; per definition they are identical in all
bins.) On the left hand side we show the shape of the DM component and
on the right hand side the shape of the gas component. The mass
denotes always the total mass (DM + gas) of the object. Filled
triangles denote the mean ratio of $a_3/a_2$ and filled diamonds
$a_3/a_1$, the latter is roughly the product of the two previous. One
can clearly see that the gas distribution is much more spherical than
the DM distribution. In both cases more massive objects are more
ellipsoidal. These objects have been formed later.  With time they
will become more spherical too.

\begin{figure}[h]
\begin{center}
\includegraphics[width=12cm]{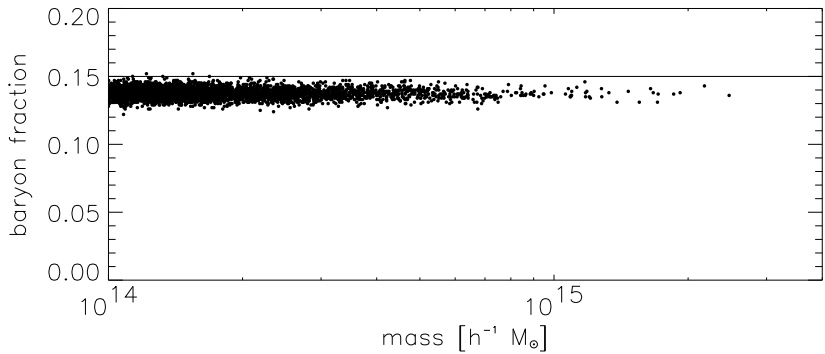}
\caption{Baryon fraction in clusters of galaxies.}
\label{GYWS:baryon}
\end{center}
\end{figure}

\begin{figure}[h]
\begin{center}
\includegraphics[width=14cm]{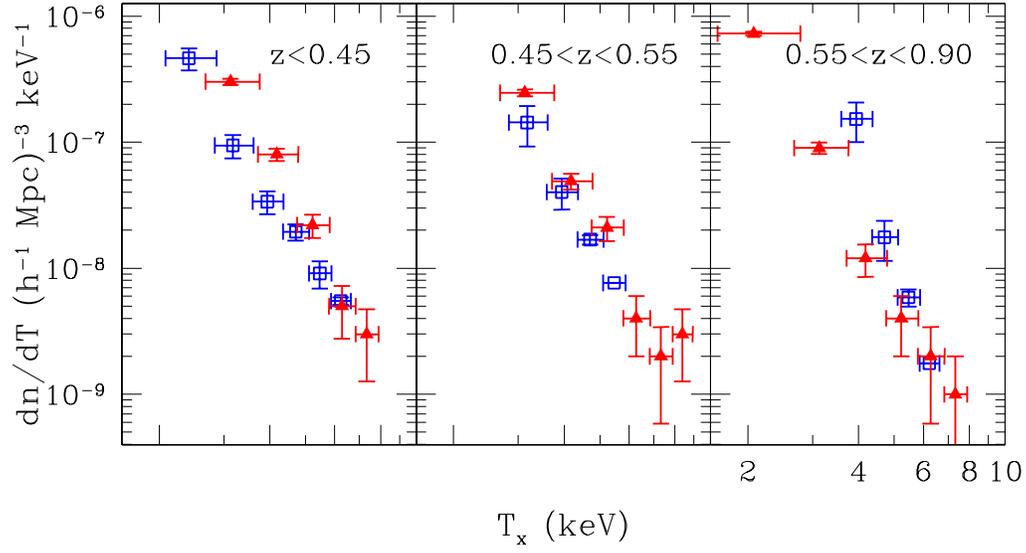}

\caption{X-ray temperature function for simulated (filled triangles)  
  and observed (open squares) CHANDRA clusters (Vikhlinin 2006) at
  different redshifts}
\label{GYWS:xraytemp}
\end{center}
\end{figure}

\section{Clusters of Galaxies}

One of the aims of the simulation was to compare the large scale
distribution of gas and DM. Since the shape of the gas distribution
and the dark matter distribution in galaxy clusters is quite different
we have determined the gas fraction in clusters within spheres
centered at the highest density peak of the cluster. The highest
density peak has been determined within the hierarchical
friends-of-friends algorithm using a linking length which corresponds
to 512 times the virial overdensity. The center of mass of this
central object has been assumed to be the center of the cluster.

In Fig.~\ref{GYWS:baryon} we show the baryon fraction in 4000
clusters with virial masses larger than $10^{14} \hMsun$. The solid
line denotes the cosmological baryon fraction of 0.15. One can clearly
see that the baryon fraction in almost all of the clusters is below
the cosmological one, the mean is about 0.14.

We have estimated the X-ray properties of the clusters found in this
simulation. We used Sutherland \& Dopita \cite{GYWS:sd} cooling
curves for primordial composition to derive X-ray emission from gas
particles. In Fig.~\ref{GYWS:xraytemp} we compare the cluster
abundance as a function of X-ray emission weighted temperature from
the MareNostrum Universe with the most recent estimates from a sample
of CHANDRA clusters \cite{GYWS:av} up to $z\sim 1$.
The agreement is remarkable at lower $z$, while we find a smaller
number of clusters at high $z$ possibly due to resolution effects.
Evolution of scaling relations have also been studied for our cluster
sample.  No significant evolution with redshift for the slope (range
from 1.8 -2.1) of the $L_x-T_x$ is found. 

\begin{figure}[!]
\begin{center}
\includegraphics[width=6cm]{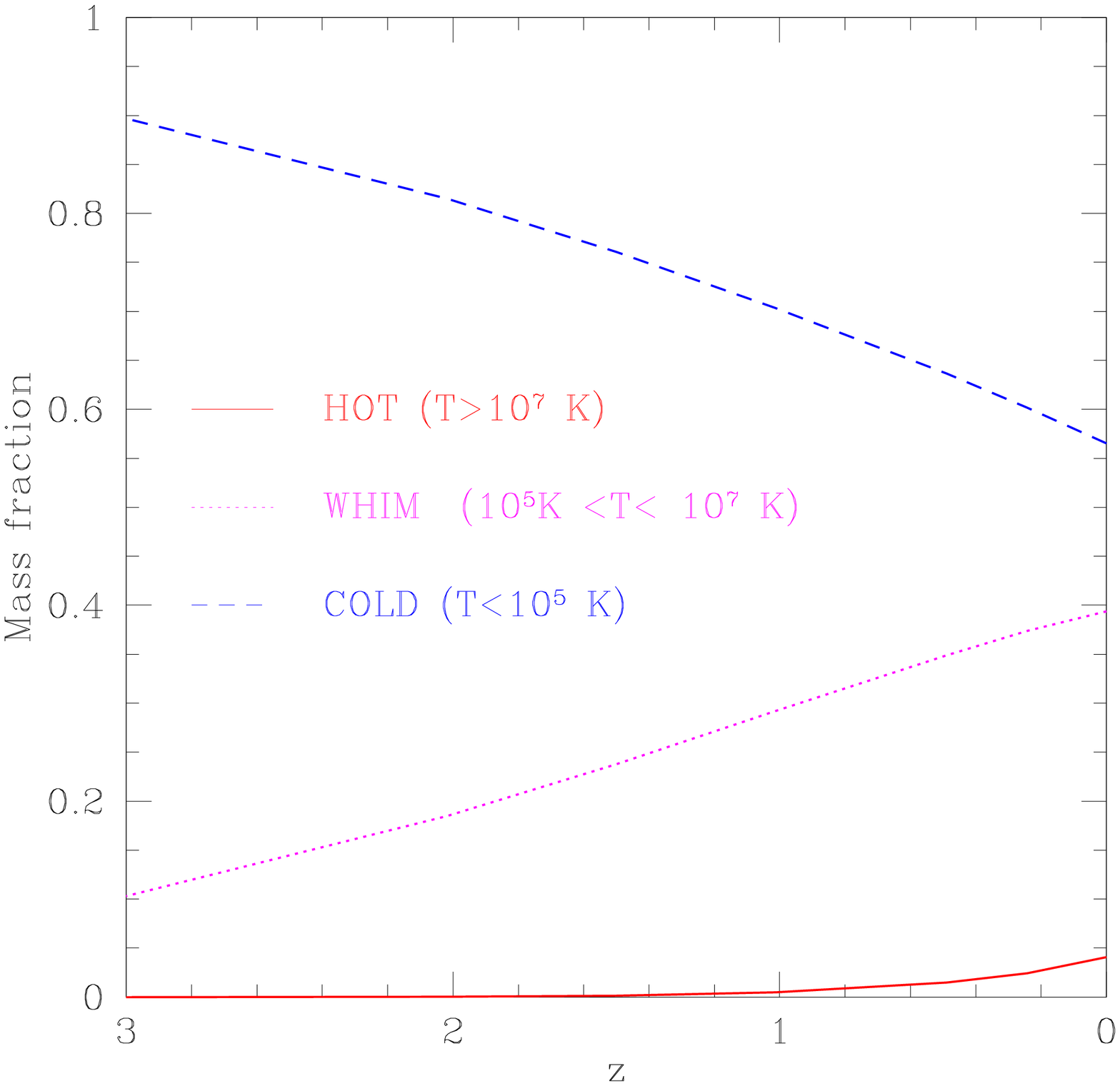}
\includegraphics[width=6cm]{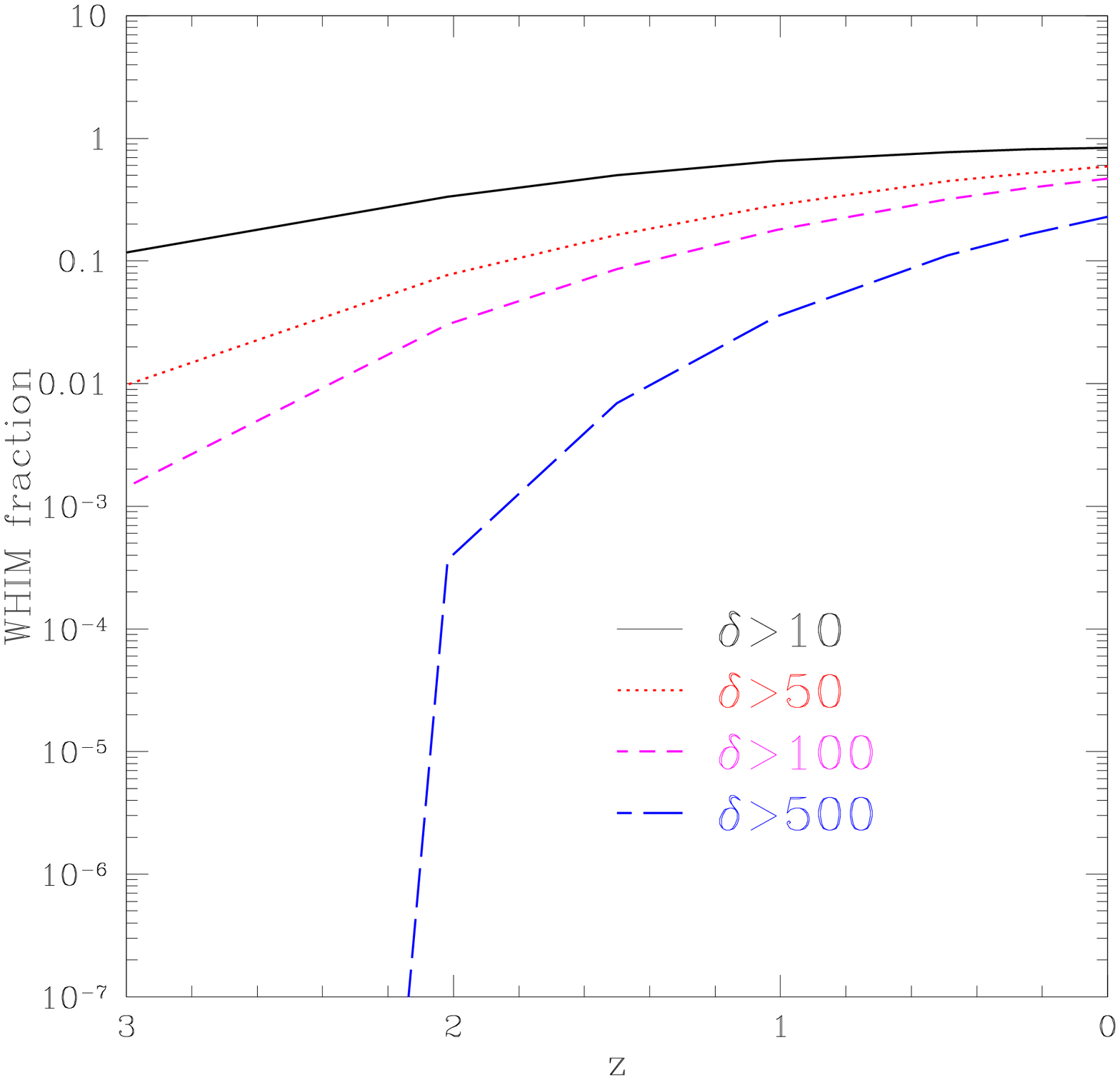}
\caption{Left: Redshift evolution of the different baryon
  phases. Right: Evolution of WHIM at different overdensities.}
\label{bfrac}
\end{center}
\end{figure}

\section{Evolution of the Baryon Distribution at Large Scales}

The large number of gas particles of this simulation allows us to
compute accurately the redshift evolution of the baryonic phase space
$(T-\rho_{gas})$.  In Fig~\ref{bfrac} we show the evolution of the
relative fractions of the different baryon components: HOT, COLD and
WARM-HOT gas.  At present, the amount of gas in the form of WHIM is of
the order of 40\%.  In the right panel of the same figure, we plot the
evolution of the WHIM component above different comoving
overdensities.  The relative fraction increases rapidly at early
redshift and becomes stable when the universe is dominated by
exponential expansion ($z < 0.5$ ). This is in good agreement with the
assumption that baryons follow a lognormal density probability
function \cite{GYWS:am}.  At $\delta > 50$ the WHIM are within
virialized objects. We see that this amounts to 60\% of the total WHIM
at present.  Therefore, we deduce that almost 40 \% of baryons live in
the WHIM phase in regions below viral overdensites, i.e. filaments and
voids.

\section{Baryonic Oscillations}

The acoustic oscillations of the primordial plasma before
recombination are not only reflected in the prominent peak structure
of the CMB power spectrum but also imprinted in the power spectrum of
the dark matter distribution, though the effect is much smaller ($\sim
5\%$) than for the CMB photons. The physical scale of these tiny
wiggles is determined by the matter and baryon densities, which will
be measured with high accuracy by the upcoming Planck mission. Knowing
the physical size the baryonic wiggles then provide a standard ruler.
If one is able to measure accurately enough the observed size of this
standard ruler transverse and parallel to the line of sight at
different redshifts, one is able to determine the angular diameter
distance $D_A(z)$ and the Hubble parameter $H(z)$ for different
redshifts and by this can constrain the equation of state parameter
$w$ of the dark energy. For details see e.g. Seo \& Eisenstein
\cite{GYWS:seo}.

\begin{figure}[t]
\begin{center}
\includegraphics[width=5.5cm, angle=-90]{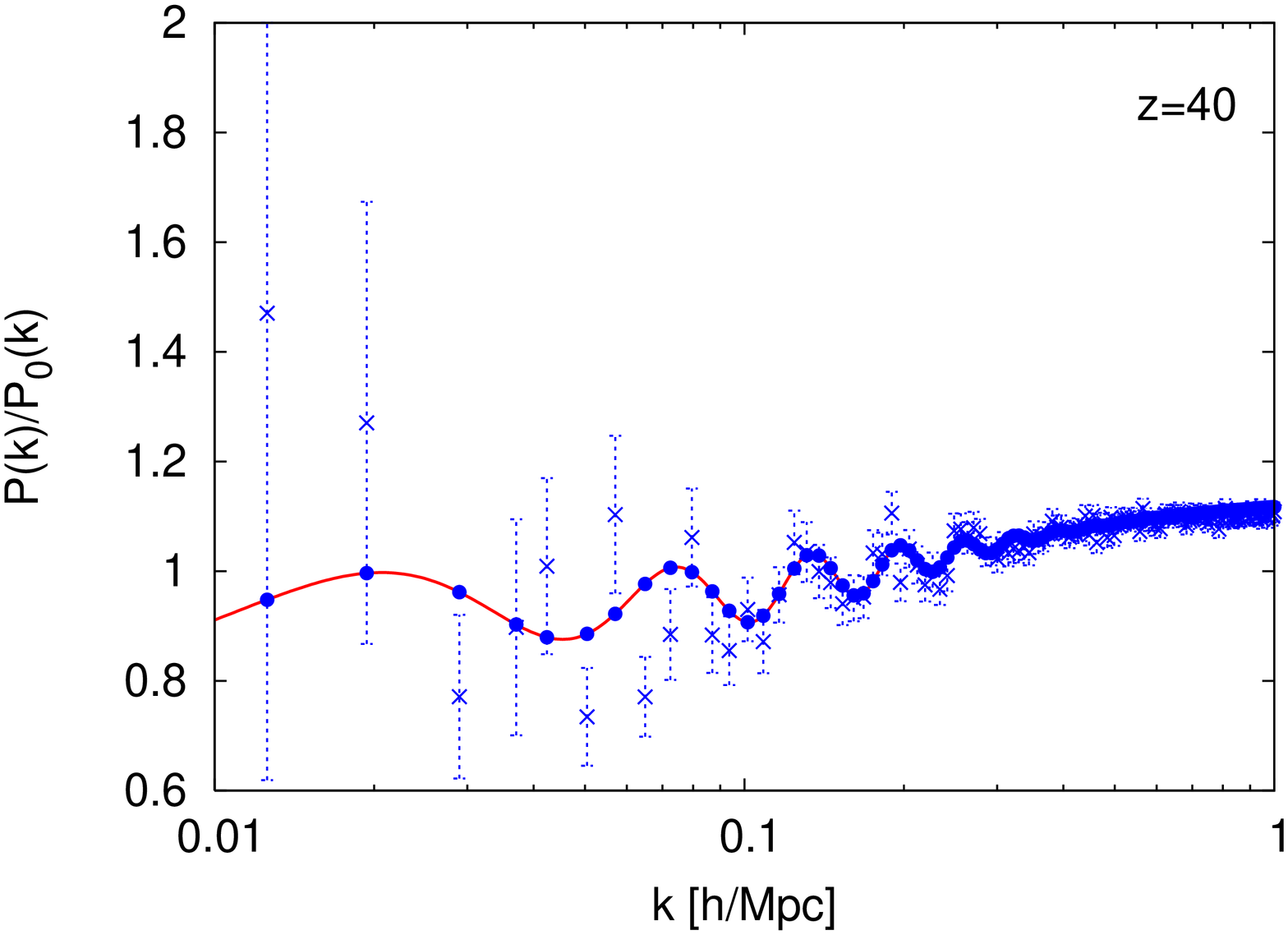}
\includegraphics[width=5.5cm, angle=-90]{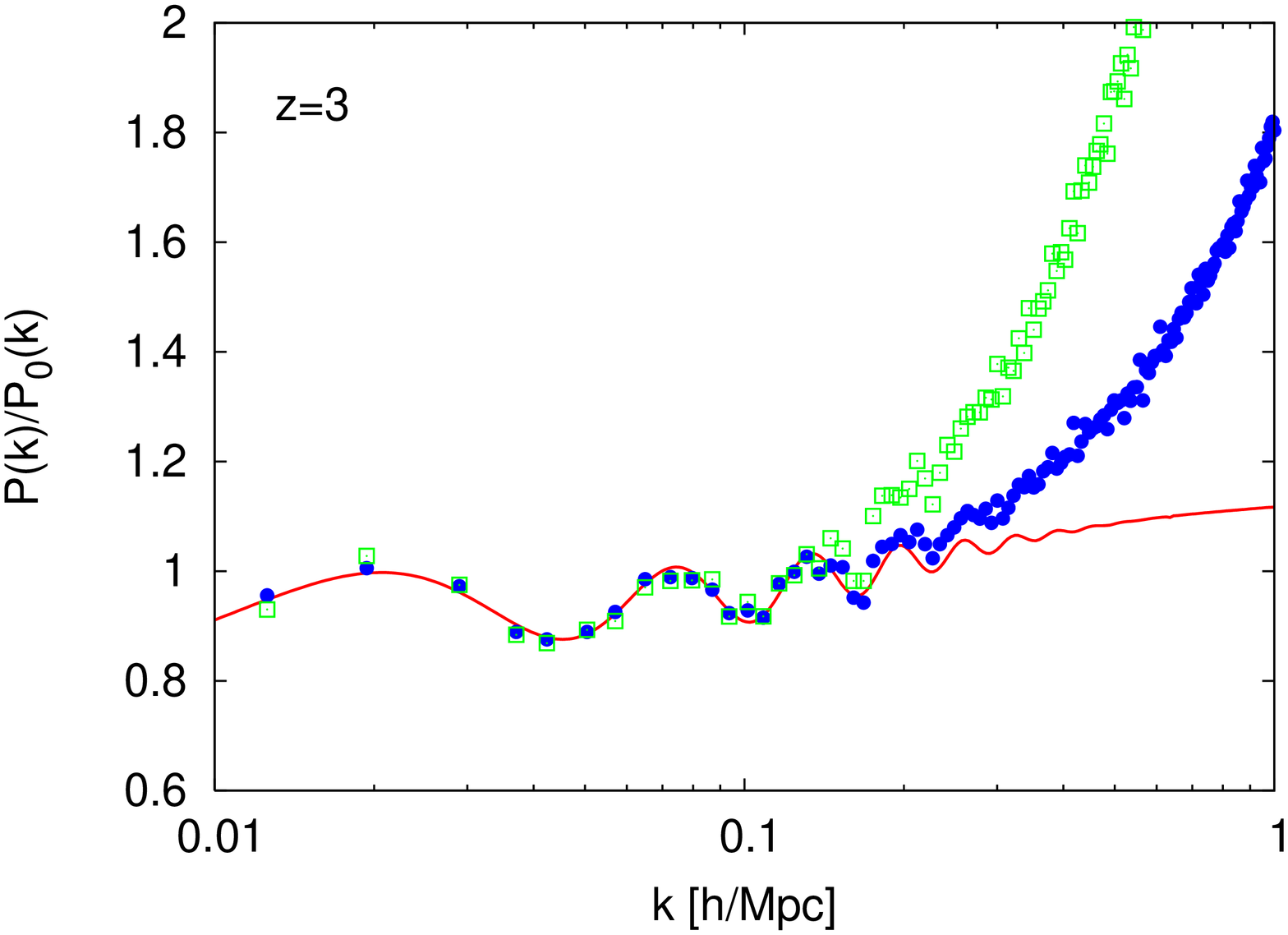}
\includegraphics[width=5.5cm, angle=-90]{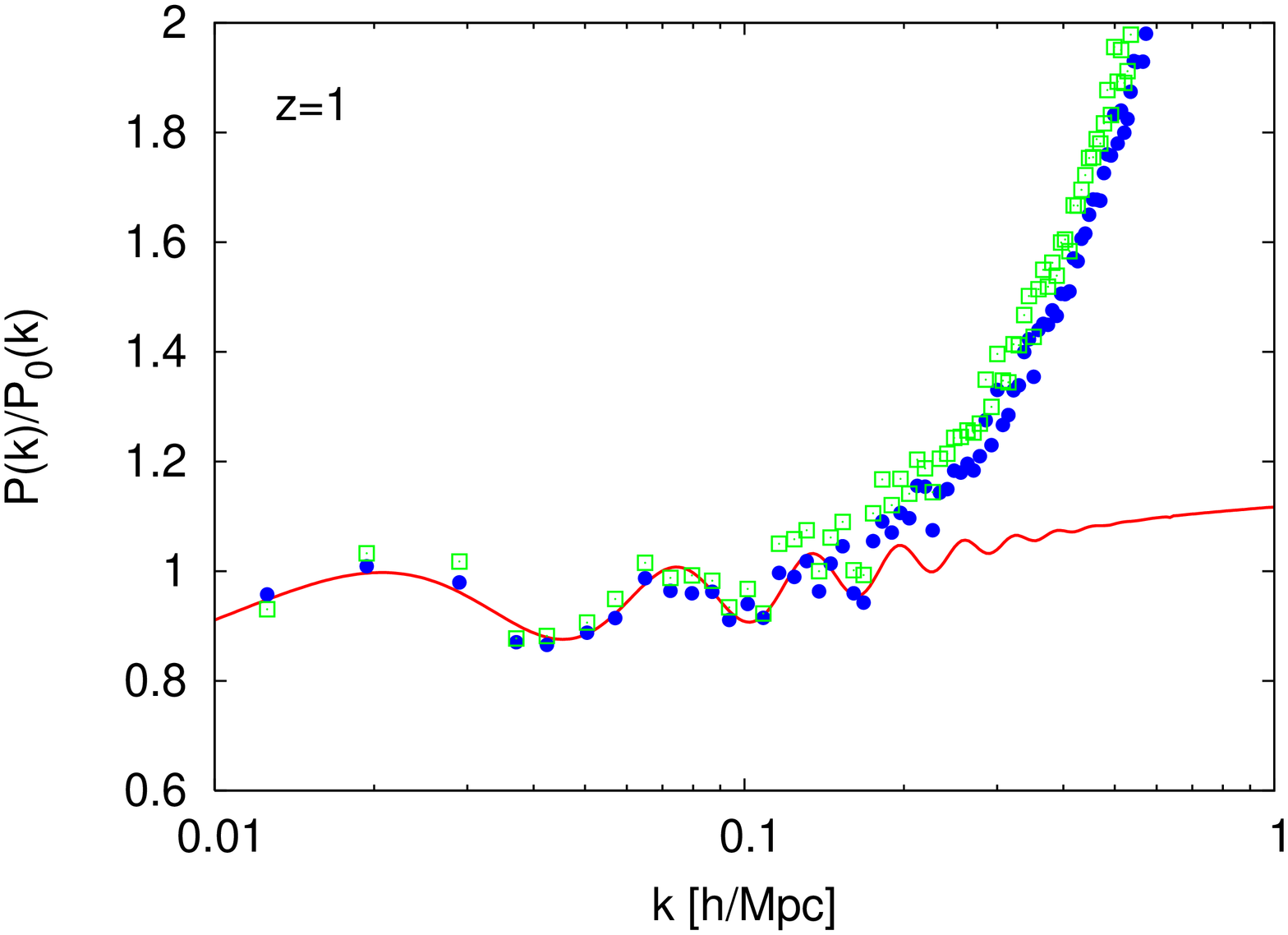}
\includegraphics[width=5.5cm, angle=-90]{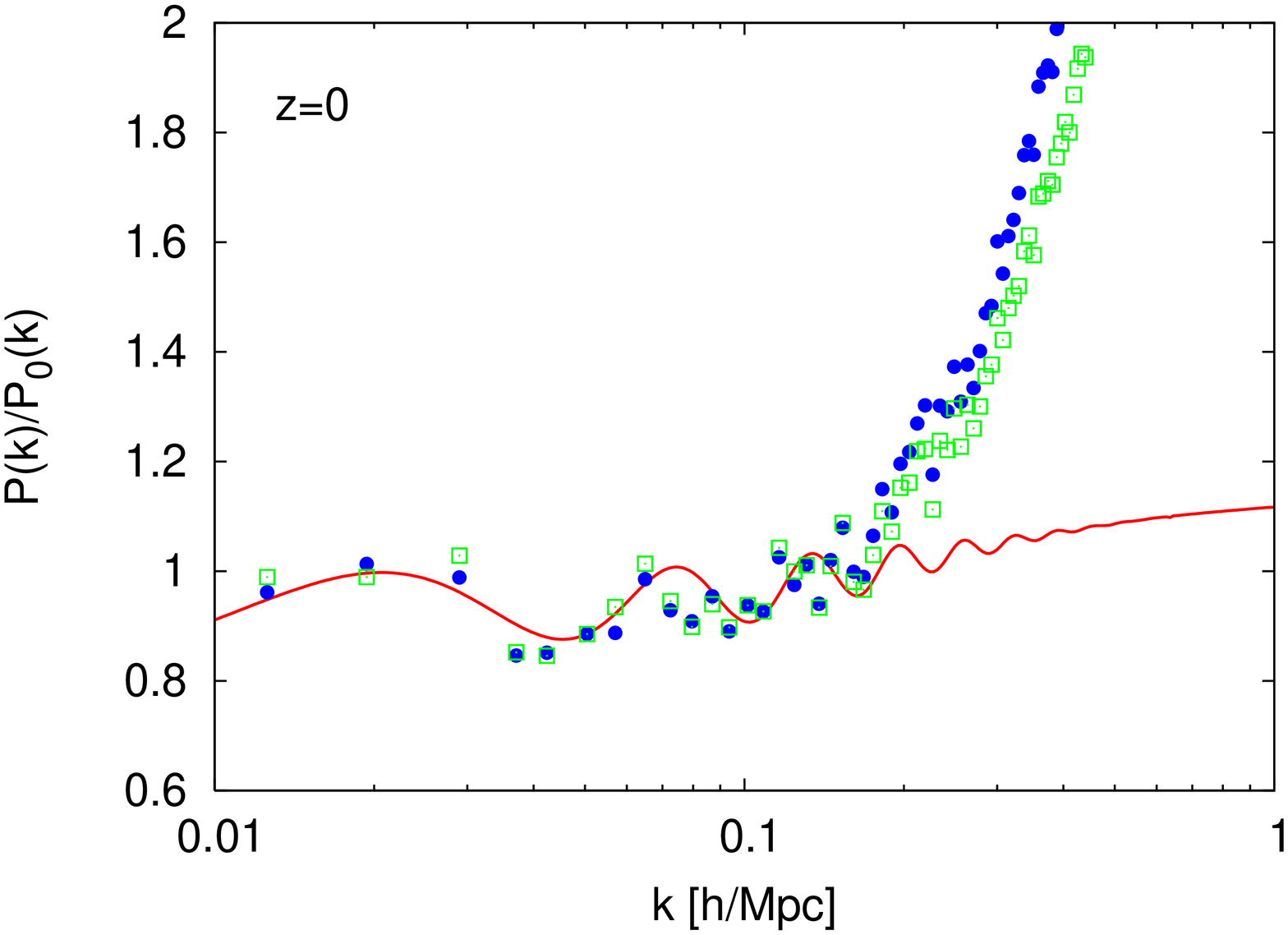}

\caption{Real-space power spectra of the dark matter (filled circles) 
  and of the halo (open squares) distributions at different redshifts.
  In all plots a linearly evolved power spectrum without baryons is
  divided out to make the baryonic oscillations more pronounced. At
  the upper left the actual realization of the initial power spectrum
  at redshift 40 is presented (crosses with error bars). The
  deviations from the mean power spectrum (solid line) due to the
  finite number of modes are corrected in all the other plots. In
  addition the halo power spectra are divided by $b^2$, where the bias
  factor $b$ is $2.78$ at $z=3$, 1.28 at $z=1$ and 0.94 at $z=0$.}
\label{GYWS:bao}
\end{center}
\end{figure}

With this simulation we were able to study qualitatively how the
non-linear evolution and the effect of galaxy bias distort the
baryonic oscillations. In Fig.~\ref{GYWS:bao} we show the real-space
power spectra of the dark matter and FOF-cluster distribution. To
enhance the visibility of the oscillations all power spectra have been
divided by a linearly evolved power spectrum with the same
cosmological parameters except for the baryon density being zero. To
this end we used the fitting formula provided by Eisenstein \& Hu
\cite{GYWS:ehu}. In the upper left panel of Fig.~\ref{GYWS:bao}
the realization (crosses) of the initial power spectrum (solid line)
used to set up the initial conditions for the simulation at redshift
40 is shown. The deviations from the desired input power spectrum are
due to the few modes available for the largest scales of the
simulation box. These deviations make it impossible to detect the
first peaks in a survey with a volume comparable to that of this
simulation. To study nevertheless the effects of non-linear evolution
and galaxy bias on the baryonic oscillation we correct all power
spectra by these initial deviations.

At $z=3$ the first three peaks ($k\sim 0.7$, $k\sim 1.3$ and $k\sim
2.0\, h/{\rm Mpc}$) of the baryonic oscillations are clearly visible
in the dark matter power spectrum. In the cluster power spectrum only
the first two remain undistorted. For lower redshifts the non-linear
evolution erases more and more the oscillation features. At $z=0$ one
can hardly see any oscillation feature. Even the first peak is
strongly distorted although it still does not lie in the non-linear
regime.

We conclude that at high redshift ($z\sim 3$) the baryonic oscillation
are conserved well enough even in the galaxy distribution to measure
their scale accurately enough to constrain the equation of state
parameter $w$ of the dark energy. The main problem will be to observe
a high number of galaxies in a very large volume to reduce both cosmic
variance and shot noise. For low redshifts it will be hard to measure
the sound horizon with the required accuracy unless one understands
the non-linear evolution well enough to reconstruct the baryonic
oscillation. One tool to learn more about this are cosmological
simulations.

\section*{Acknowledgments}
We would like to thank the Barcelona Supercomputer Center for allowing
us to run the simulation described above during the testing period of
MareNostrum. The analysis of this simulation has been done on NIC
J\"ulich. We also thank Acciones Integradas Hispano-Alemanas for
supporting our collaboration.

\end{document}